\documentclass[aps, amsmath, amsfonts]{revtex4}
\usepackage{graphicx}

\DeclareMathOperator{\tr}{Tr}

\newcommand{\nn}{\nonumber} 
\newcommand{\eq}[1]{eq.~(\ref{#1})}
\newcommand{\Eq}[1]{Eq.~(\ref{#1})}

\newcommand{\sla}[1]{#1 \! \! \! \slash}

\renewcommand{\bar}[1]{\overline{#1}}

\begin{document}

\preprint{ \hbox{MIT-CTP 3568} 
           \hbox{hep-ph/} }

\title{Constraints on widths of mixed pentaquark multiplets} 

\author{Vivek Mohta}
\affiliation{Center for Theoretical Physics and \\ Laboratory for Nuclear Science, \\ Massachusetts Institute of Technology \\ Cambridge, MA 02139\\  \\ Department of Mathematics, \\ Harvard University \\ Cambridge, MA 02138}

\begin{abstract}
We determine constraints on the partial widths of mixed pentaquark multiplets in the framework of heavy baryon chiral perturbation theory (HB$\chi$PT). The partial widths satisfy a GMO-type relation at leading order in HB$\chi$PT, for arbitrary mixing. The widths of $N(1440)$, N(1710), and $\Theta(1540)$ are not consistent with ideal mixing, $\theta_{N} = 35.3^{\circ}$, but are consistent with $\theta_{N} \lesssim 25^{\circ}$. Furthermore, there are parameter values in HB$\chi$PT that produce such a mixing angle while allowing the identification of the mass spectrum above. As an alternative to non-ideal mixing, we also suggest reasons for giving up on $N(1440)$ as a pure pentaquark state.
\end{abstract}

\maketitle

\section{Introduction}
Several experiments in the last year have found evidence for the exotic hadronic resonance $\Theta^+(1540)$. Evidence to the contrary has also appeared in recent months, but there is no final word yet on the story of pentaquarks \cite{Trilling}. In a previous paper, we considered properties of an anti-decuplet of pentaquarks in the framework of chiral perturbation theory \cite{Mohta}. We calculated the width at leading order (LO) and calculated the NNLO corrections to the masses. However, we postponed the inclusion of other multiplets of pentaquarks and ignored mixing. These are taken up in this paper.

States of four quarks and an antiquark can come in several different $SU(3)_{f}$ multiplets: $\mathbf{1, 8, \bar{10}, 10, 27, 35}$. Different models allow different multiplets. The smallest multiplet including an explicitly exotic state is $\mathbf{\bar{10}}$. It appears as a summand, along with an octet, in the product $\mathbf{\bar{6}}_{4q} \otimes \mathbf{\bar{3}}_{\bar{q}}$. In other words, an anti-decuplet is inevitably accompanied by an octet \cite{Pras}. In the Jaffe-Wilczek (JW) diquark model, the octet and anti-decuplet are degenerate \cite{JaffeWilczek}. The same should be true in the Karliner-Lipkin diquark-triquark model \cite{KarlinerLipkin}. Other treatments allow different masses for the two multiplets \cite{Diakonov, ManoharExotic}.

Decays of the pentaquark candidate $\Theta$ into nucleon and kaon have a very small width. This is particularly suprising because no quark pair creation is required as in the decays of ordinary baryons \cite{JaffeJain}. Epxerimental estimates of its widths vary from less than one MeV to a few MeV. Even the largest estimates suggest that the wavefunctions of pentaquarks are considerably different than that of nucleons \cite{JaffeWilczek}. In the model in \cite{Hong}, the large suppression has been more explicitly described by a barrier to the tunneling of a quark that allows the decay of the state $\Theta^{+}$ into $N K$. Suppression in several quark models is discussed in \cite{Maltman}. 

The hypothesis of small overlap and the mass difference between the anti-decuplet of pentaquarks and the octet of baryons suggest that the do not substantially mix the members of these multiplets. However, if there is also an octet of {\it pentaquarks} with a similar mass, it is quite possible that the quark masses mix the two multiplets of pentaquarks. In fact, in the JW, the pentaquark multiplets are degenerate in the absence of $SU(3)_f$ symmetry violations and mix ideally due to the symmetry-violating quark masses, like the octet and singlet of light vector mesons.

Phenomenological constraints on, and difficulties with, the identification of the hadronic spectrum proposed by Jaffe and Wilczek have been discussed before. Ideal mixing fits the {\it masses} of the recently discovered pentaquark candidates $\Theta(1540)$ and $\Phi(1860)$ and the known resonances $N(1440)$,  $\Sigma(1660)$, and $N(1710)$. However, it is difficult to accomodate the order of magnitude larger {\it width} of the $N(1440)$ \cite{JaffeWilczek, Cohen}. In \cite{Diakonov}, it is argued that  $N(1710)$, $\Lambda(1800)$, $\Sigma(1880)$, and $\Xi(1950)$ fill out an octet since their masses satisfy the Gell-Mann-Okubo relation so well. It is also suggested that $N(1440)$, $\Lambda(1600)$, $\Sigma(1660)$, and $\Xi(1690)$, which do not satisfy the GMO relation to the same degree, are candidates for mixing with the anti-decuplet. On the other hand, in \cite{Glozman} it is argued that $N(1440)$, $\Lambda(1600)$, $\Sigma(1660)$, and $\Delta(1600)$ fill out a $\mathbf{56}$ of spin-flavor $SU(6)$ based on their similar widths and based on a successful constituent quark model with Goldstone boson exchange \cite{Glozman2}. The PDG tentatively classifies $N(1440)$, $\Lambda(1600)$, and $\Sigma(1660)$ as members of a radially excited $\mathbf{56}$ of $SU(6)$, but it does not suggest an $SU(6)$ classification for $\Delta(1600)$ \cite{PDG}.

In this paper, we consider the phenomenological implications of mixing of approximately degenerate pentaquark multiplets in the framework of heavy chiral perturbation theory (HB$\chi$PT). Chiral perturbation theory was first applied to pentaquarks to study production mechanisms \cite{Ko}. HB$\chi$PT was later used to provide phenomenological constraints on the spin and parity of pentaquarks \cite{Mehen}. Flavor symmetry was also used to relate decay amplitudes of pentaquarks into baryon-meson final states, for each allowed exotic multiplet \cite{Grinstein, Golbeck}.

We use HB$\chi$PT to calculate masses and partial widths for mixed pentaquark multiplets. The chiral expansion allows us to generalize Cohen's inequality to a GMO-type relation satisfied by the {\it reduced partial widths} at leading order in HB$\chi$PT for arbitrary mixing. We find that the ideal mixing in the JW model and the identification of $N(1440)$, N(1710), and $\Theta(1540)$ proposed is not consistent with the width data. Non-ideal mixing satisfying $\theta_{N} \lesssim 25^{\circ}$ is however consistent with the width data.

As an alternative to changing the mixing angle, we suggest changing the assignment of $N(1440)$. This resonance is very likely not a pure pentaquark state based on the large partial width for $\Delta(1600) \to N(1440) \pi$. Although we treat the JW model in detail, our analysis can be applied to other models for the an anti-decuplet and octet for which there are a sufficient number of predictions to match on to. In fact, the analysis is valid even if the octet is made up of ordinary baryons rather than pentaquarks.

The paper is organized as follows. In section \ref{mixing}, we describe the mixing in terms of the couplings in the Lagrangian and calculate partial widths for the mass eigenstates to decay into a nucleon and a meson. In section \ref{disc} we discuss the results of other authors and find the constraints our results impose on the identification of existing resonances. In section \ref{conc} we have some concluding remarks.

\section{Mixing in $\chi$PT} \label{mixing}

\begin{figure}[htdp]
\caption{ \label{weight} Overlayed weight diagrams for octet and anti-decuplet of $SU(3)_{f}$}
\begin{center}
\includegraphics{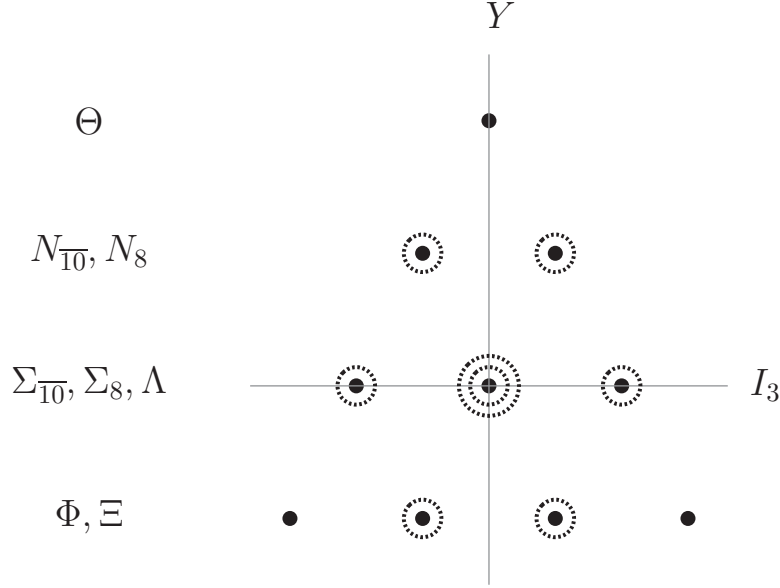}
\end{center}
\end{figure}

We briefly review our field definitions and conventions \cite{Mohta} and then construct the desired Lagrangian.  The pseudo-Goldstone bosons are described by the field $\xi = \exp (i \pi /f)$ where $f = 131 \, \textrm{MeV}$ is the pion decay constant and the field $\pi$ is 
\begin{equation}\label{pifield}
\pi = \begin{bmatrix}
\frac{1}{\sqrt{2}}\pi^0 + \frac{1}{\sqrt{6}}\eta & \pi^+ & K^+ \\
\pi^- & -\frac{1}{\sqrt{2}}\pi^0 +\frac{1}{\sqrt{6}}\eta & K^0 \\
K^- & \bar{K}{}^0 & -\frac{2}{\sqrt{6}}\eta \\
\end{bmatrix}.
\end{equation}
The ground state octet of baryon fields is
\begin{equation}\label{bfield}
B = \begin{bmatrix}
\frac{1}{\sqrt{2}}\Sigma^0 + \frac{1}{\sqrt{6}}\Lambda & \Sigma^+ & p \\
\Sigma^- & -\frac{1}{\sqrt{2}}\Sigma^0 + \frac{1}{\sqrt{6}}\Lambda & n \\
\Xi^- & \Xi^0 & \frac{2}{\sqrt{6}}\Lambda\\
\end{bmatrix}.
\end{equation}
The anti-decuplet of pentaquark fields is
\begin{align}\label{pfield}
P_{333} &= \Theta^{+}  \\
P_{133} = \frac{1}{\sqrt{3}} \, N^0_{\bar{10}}  &\qquad 
P_{233} = \frac{1}{\sqrt{3}} \, N^+_{\bar{10}} \nn \\
P_{113} = \frac{1}{\sqrt{3}} \, \Sigma^{-}_{\bar{10}}\qquad 
P_{123} &= \frac{1}{\sqrt{6}}\Sigma^{0}_{\bar{10}}\qquad 
P_{223} = \frac{1}{\sqrt{3}} \, \Sigma^{+}_{\bar{10}}   \nn \\
P_{111} =  \Phi^{--} \qquad 
P_{112} = \frac{1}{\sqrt{3}}  \, \Phi^{-}  &\qquad 
P_{122} = \frac{1}{\sqrt{3}} \, \Phi^{0} \qquad 
P_{222} =  \Phi^{+}\, . \nn
\end{align}
In the literature, $\Phi$ is sometimes called $\Xi_{\bar{10}}$. The octet of pentaquark fields is
\begin{equation}\label{ofield}
O = \begin{bmatrix}
\frac{1}{\sqrt{2}}\Sigma^0_{8} + \frac{1}{\sqrt{6}}\Lambda_{8} & \Sigma^+_{8} & N^{+}_{8} \\
\Sigma^-_{8} & -\frac{1}{\sqrt{2}}\Sigma^0_{8} + \frac{1}{\sqrt{6}}\Lambda_{8} & N^{0}_{8} \\
\Xi^-_{8} & \Xi^0_{8} & \frac{2}{\sqrt{6}}\Lambda_{8}\\
\end{bmatrix}.
\end{equation}
In the analysis of mixing, it will be helpful to have overlayed weight diagrams for these multiplets, as in 
figure \ref{weight}.

It is convenient to trade in the $\xi$ field and its derivative for two fields with simple transformation properties:
\begin{align}
\label{amu} A_\mu & = \frac i 2 \left(\xi\partial_\mu \xi^\dagger - \xi^\dagger \partial_\mu \xi\right) = \frac{1}{f}\partial_\mu \pi - \frac{1}{6 f^3}\left[\left[\partial_\mu\pi, \pi \right],\pi\right] + \ldots \\
\label{vmu} V_\mu & = \frac 1 2 \left(\xi\partial_\mu \xi^\dagger + \xi^\dagger \partial_\mu \xi\right) = \frac{1}{2f^2}\left[\pi,\partial_\mu \pi \right] + \ldots
\end{align}
In QCD, the quark mass $M_q$ explicitly breaks chiral symmetry. However, if $M_q$ is assumed to transform as $ M_q \rightarrow L M_q R^\dagger $ then the QCD Lagrangian is invariant under chiral symmetry. In order to have the same symmetry breaking pattern in chiral perturbation theory, we construct invariant terms involving an $M_q$ which transforms in this way. Furthermore, it is convenient to trade in $M_q$ for the combination of $M_q^\xi = (\xi^\dagger M_q \xi^\dagger + \xi M_q^\dagger \xi)$ which transforms like the other fields.

A covariant derivative can be defined for each matter field by
\begin{align} \label{covder}
D_\mu B &= \partial_\mu B + \left[V_\mu,B\right] \\
D_\mu O &= \partial_\mu O + \left[V_\mu,O\right] \nn \\
\left(D_\mu P\right)_{ijk} &= \partial_\mu P_{ijk} - P_{ljk} V_\mu^l{}_i - P_{ilk} V_\mu^l{}_j - P_{ijl} V_\mu^l{}_k \nn
\end{align}
As usual, the covariant derivatives of the fields transform like the fields themselves. Under parity, the fields $B$, $V_\mu$, and $M_q^\xi$ are even, while the field $A_\mu$ is odd. In this paper, we will assume that $P$ and $O$ have $J^{P} = \frac{1}{2}^{+}$.

With three multiplets of baryon fields with different masses, an independent heavy fermion field redefinition \cite{Jenkins} of each results in simple kinetic terms for all fields but introduces a phase in the interaction terms. An equivalent possibility, to keep the interaction terms simple, is to use the same field redefinition for all three fields,
\begin{align} \label{hqfield}
P_v(x) &= \frac{1}{2} \left(1 + \sla{v} \right) e^{i m_P v \cdot x} P(x) \\
O_v(x) &= \frac{1}{2} \left(1 + \sla{v} \right) e^{i m_P v \cdot x} O(x) \nn \\
B_v(x) &= \frac{1}{2} \left(1 + \sla{v} \right) e^{i m_P v \cdot x} B(x) \nn
\end{align}
where $m_P$ is the mass of the anti-decuplet. Doing so introduces the constants $\Delta_{PO} = m_P - m_O$ and $\Delta_{PB} = m_P - m_B$ in the kinetic terms for $O_{v}$ and $B_{v}$. We adopt this second approach.

The LO terms in the chiral Lagrangian involving only the anti-decuplet are \cite{Mehen}
\begin{equation}
    \label{lagp} \mathcal{L}^{(1)}_P = \bar{P}_v i v \cdot \mathcal{D} P_v + 2 \mathcal{H}_{P} \bar{P}_v S_v \cdot A P_v.
\end{equation}    
The NLO terms involving only anti-decuplet and having quark masses are \cite{Mohta}
\begin{equation}    
    \label{lagpmass} \mathcal{L}_P^{(2)} = \sigma_P \frac{B_0}{4 \pi f} (\tr M_q^\xi) \bar{P}_v P_v + b_{\mathcal{H}} \frac{B_0}{4 \pi f} \bar{P}_v M_q^\xi P_v + \ldots 
\end{equation}
The flavor indices have been suppressed. Details can be found in \cite{Mohta}. We have included a factor of $B_{0}/\Lambda$ in the quark mass terms where $\Lambda$ is the cutoff for the effective theory. This convention trades in a choice of $m_{q}$ for the observable $m_{\pi}^{2}$ and a choice of $\Lambda$, and makes the power counting of terms explicit. As in \cite{Mohta}, we choose $\Lambda$ to be $4 \pi f$.

The terms involving only the octet of pentaquarks are the same as for an octet of ordinary baryons,
\begin{align}
    \label{lago} \mathcal{ L}^{(1)}_{O} &=  \tr \left(\bar{O}_v \left(i v \cdot \mathcal{D} + \Delta_{PO} \right) O_v \right) \\ &+ 2 D_{O} \tr \left(\bar{O}_v S^\mu_v \left\{A_\mu,O_v\right\}\right) + 2 F_{O}  \tr \left(\bar{O}_v S^\mu_v \left[A_\mu,O_v\right] \right) \nn \\ 
   \label{lagomass}  \mathcal{ L}^{(2)}_{O} &=  \sigma_O \frac{B_0}{4 \pi f} \tr M_q^\xi \tr \left( \bar{O}_v O_v \right) \\ &+ b^{D}_{O} \frac{B_0}{4 \pi f} \tr \left(\bar{O}_v  \left\{M_q^\xi, O_v\right\}\right) + b^{F}_{O} \frac{B_0}{4 \pi f} \tr \left(\bar{O}_v \left[M_q^\xi, O_v\right] \right) \ldots \nn
\end{align}
As in \cite{Mohta}, the same field transformation has been performed on all matter fields, and $\Delta_{PO} = m_{P} - m_{O}$ is the residual mass difference between the anti-decuplet and octet in the chiral limit. 
The terms in the Lagrangian involving both the anti-decuplet and octet are
\begin{align}
   \label{lagpo} \mathcal{L}^{(1)}_{PO} &= 2 \mathcal{C}_{P O} \left(\bar{P}_v S_v \cdot A O_v + \bar{O}_v S_v \cdot A P_v \right) \\
   \label{lagpomass} \mathcal{L}^{(2)}_{PO} &= b_{P O}  \frac{B_0}{4 \pi f} \left(\bar{P}_v M_q^\xi O_v + \bar{O}_v M_q^\xi P_v \right)+ \ldots 
\end{align}
We won't have much use for $\mathcal{L}^{(1)}_{PO}$ in this paper, but $\mathcal{L}^{(2)}_{PO} $ is responsible for mixing between the multiplets.

Flavor symmetry alone limits to six, the parameters needed to fit the masses of an octet and anti-decuplet mixed by a quark mass perturbation. Two parameters set the overall masses of the multiplets, two parameters describe the shift in the octet, one parameter describes the shift in the anti-decuplet, and one parameter describes the mixing. In HB$\chi$PT, the six parameters are
\begin{equation} m_{P} - 2 \sigma_P \frac{m_{K}^{2}}{4 \pi f}, \quad m_{O} - 2 \sigma_O \frac{m_{K}^{2}}{4 \pi f}, \quad b^{D}_{O} \frac{m_{K}^{2}}{4 \pi f}, \quad b^{F}_{O} \frac{m_{K}^{2}}{4 \pi f}, \quad  b_{\mathcal{H}} \frac{m_{K}^{2}}{4 \pi f}, \quad \textrm{and} \quad b_{P O}  \frac{m_{K}^{2}}{4 \pi f}.
\end{equation}
For simplicity, we set $\hat{m}= m_{u} = m_{d}$  to zero while keeping $m_{s}$ nonzero. A nonzero $\hat{m}$ would not introduce any new parameters. 

Although it seems that we have eight parameters, the parameters $\sigma_{P}$ and $\sigma_{O}$ shift all the masses in a multiplet by an equal amount and hence can be absorbed in the parameters $m_{P}$ and $m_{O}$ respectively, when fitting physical masses. They take on a more important role in chiral extrapolation and in fitting scattering data.

From figure \ref{weight}, we conclude that mixing is limited to two isospin multiplets, $N$ and $\Sigma$, in each pentaquark multiplet. The octet has no particle with the quantum numbers of $\Theta$. The strangeness $S=-2$ multiplets $\Xi$ and $\Phi$ have isospin $\frac{1}{2}$ and $\frac{3}{2}$ respectively and therefore do not mix. The isospin singlet $\Lambda$ cannot mix with any field for the same reason.

The mixing calculations for the $N$ isospin multiplets are done explicitly below. The masses and partial widths for the remaining multiplets are displayed in Table \ref{fulltable} in section \ref{disc}. We start by noting that the mass term in the Hamiltonian for the $N$ multiplets has the form
\begin{equation} \label{ham}\
\mathcal{H}_{m} = {\begin{bmatrix} \bar{N}_{\bar{10}} \\  \bar{N}_{8} \end{bmatrix}}^{T}
\begin{bmatrix}
\, x \,  & \, y \, \\
\, y \, & \, z \,
\end{bmatrix}
{\begin{bmatrix} N_{\bar{10}} \\  N_{8} \end{bmatrix}}.
\end{equation}
The isospin multiplet $\bar{N}$, for both $\bar{10}$ and $8$, is defined by $\bar{N} = ( - \bar{N^{0}}, \bar{N^{+}} )$. The quantities $x$, $y$, and $z$ in the mass matrix are determined by the chiral Lagrangian in \eq{lagpmass}, \eq{lagomass}, and \eq{lagpomass}:
\begin{equation} \label{massmatrix}
\begin{bmatrix}
\, x \,  & \, y \, \\
\, y \, & \, z \,
\end{bmatrix}
= \frac{m_{K}^{2}}{4 \pi f}
\begin{bmatrix}
- \frac{4}{3} b_{\mathcal{H}} - 2 \sigma_{P} & \quad  - \frac{2}{\sqrt{3}} b_{PO}  \\
 - \frac{2}{\sqrt{3}} b_{PO}  & \quad - 2 b_O^{D}+ 2 b_O^{F} - \bar{\Delta}_{PO} - 2 \sigma_{P} 
\end{bmatrix}.
\end{equation}
Here $\bar{\Delta}_{PO}= \frac{4 \pi f}{m_{K}^{2}}[(m_{P} - 2 \sigma_P \frac{m_{K}^{2}}{4 \pi f} ) - ( m_{O} - 2 \sigma_O \frac{m_{K}^{2}}{4 \pi f})]$, or in words, $\bar{\Delta}_{PO}$ is the difference in the overall masses of the multiplets in units of $m_{K}^{2}/ 4 \pi f$. The mass matrix can be diagonalized by a change of basis of the field operators,
\begin{equation} \label{massdiagonal}
\mathcal{H}_{m}={\begin{bmatrix} \bar{N}_1 \\  \bar{N}_2 \end{bmatrix}}^{T}
\begin{bmatrix}
\left(\frac{x+z}{2}\right) + \left(\frac{x - z}{2}\right) \left(1+ \left(\frac{2y}{x-z}\right)^{2}\right)^{\frac{1}{2}} & 0 \\
0 &  \left(\frac{x+z}{2}\right) - \left(\frac{x - z}{2}\right) \left(1+ \left(\frac{2y}{x-z}\right)^{2}\right)^{\frac{1}{2}} 
\end{bmatrix}
{\begin{bmatrix} N_{1} \\  N_{2} \end{bmatrix}}.
\end{equation}
The mixing angle is defined by the relationship between the fields $N_{1}, N_{2}$ and $N_{\bar{10}}, N_{8}$:
\begin{equation} \label{rotationmatrix}
{\begin{bmatrix} N_{1} \\  N_{2} \end{bmatrix}}
=
\begin{bmatrix}
\cos \theta& \sin \theta \\
-\sin \theta& \cos \theta
\end{bmatrix}
{\begin{bmatrix} N_{\bar{10}} \\  N_{8} \end{bmatrix}}.
\end{equation}
We have chosen this convention to agree with earlier authors \cite{Jaffe, Diakonov}. The mixing angle $\theta$ is between $-45^{\circ}$ and $45^{\circ}$ and satisfies
\begin{equation} \label{angle}
\tan{2 \theta} = \frac{2 y}{z-x} .
\end{equation}

Combining \eq{massmatrix}, \eq{massdiagonal}, and \eq{angle}, we find the masses of the states $N_{1}$ and $N_{2}$ to be
\begin{align}
m_{{N}_{1}} & = \left( m_{P}-2 \sigma_{P} \frac{m_{K}^{2}}{4 \pi f}\right)  - \frac{2}{3} b_{\mathcal{H}} \left(1 + \sec{2 \theta}\right) +  \left(- b_O^{D} +  b_O^{F} - \frac{1}{2} \bar{\Delta}_{PO}\right)  \left(1 - \sec{2 \theta}\right) \\
m_{{N}_{2}} & =  \left(m_{P}-2 \sigma_{P} \frac{m_{K}^{2}}{4 \pi f}\right)  - \frac{2}{3} b_{\mathcal{H}} \left(1 - \sec{2 \theta}\right) +  \left(- b_O^{D} +  b_O^{F} - \frac{1}{2} \bar{\Delta}_{PO}\right)  \left(1 + \sec{2 \theta} \right). 
\end{align}
and the mixing angle to satisfy
\begin{equation}
\tan{2 \theta} = \frac{- \frac{2}{\sqrt{3}} b_{PO} }{ \frac{2}{3} b_{\mathcal{H}} -  b_O^{D} +  b_O^{F} - \frac{1}{2} \bar{\Delta}_{PO}}.
\end{equation} 
When $b_{PO}$ is zero, $\theta$ is zero and the mass formulas for $m_{N_{1}}$ and $m_{N_{2}}$ reduce to the GMO formulas for $N_{\bar{10}}$ and $N_{8}$ respectively.

Having determined the mass eigenstates, we now calculate their partial widths to nucleon-pion final states. The LO terms in the chiral Lagrangian involving the pentaquark multiplets and the ground state octet of baryons are
\begin{align}
    \label{lagpb} \mathcal{L}^{(1)}_{PB} &= 2 \mathcal{C}_{P B} \left(\bar{P}_v S_v \cdot A B_v + \bar{B}_v S_v \cdot A P_v \right) \\
    \label{lagob} \mathcal{ L}^{(1)}_{OB} &= 2 D_{OB} \tr \left(\bar{O}_v S^\mu_v \left\{A_\mu,O_v\right\}\right) + 2 F_{OB}  \tr \left(\bar{O}_v S^\mu_v \left[A_\mu,O_v\right] \right).
\end{align}
The width for $\Theta$ to decay into $N K$ was calculated at LO in \cite{Mehen}: 
\begin{equation}
\Gamma_{\Theta N K} = 2 \mathcal{C}^2_{PB} \frac{m_B}{m_P} \frac{k^3}{2 \pi f^2}.  
\end{equation}

To calculate the partial widths for $N_{1}$ and $N_{2}$, we trade in the fields $N_{\bar{10}}$ and $N_{8}$ in \eq{lagpb} and \eq{lagob} using \eq{rotationmatrix}. The partial widths for $N_{1}$ and $N_{2}$ into $N \pi$ at LO are
\begin{align} \label{nwidth}
\Gamma_{N_{1} N \pi} &=\frac{3}{2} \left(\frac{1}{\sqrt{3}} \mathcal{C}_{PB} \cos{\theta} - \left(D_{OB}-F_{OB}\right) \sin{\theta} \right)^{2} \left(\frac{m_{N}}{m_{N_{1}}}\right)\frac{k_{N \pi}^{3}(m_{N_{1}})}{2 \pi f^2}
 \\
\Gamma_{N_{2} N \pi} &=\frac{3}{2} \left(\frac{1}{\sqrt{3}} \mathcal{C}_{PB} \sin{\theta} + \left(D_{OB}-F_{OB}\right)\cos{\theta} \right)^{2} \left(\frac{m_{N}}{m_{N_{2}}}\right)\frac{k_{N \pi}^{3}(m_{N_{2}})}{2 \pi f^2}.
\end{align}
As usual, the kinematics imply that the meson momentum is
\begin{equation}
k_{b a}(m) =\left[\frac{(m^{2}-(m_{b}+m_{a})^{2})(m^{2}-(m_{b}-m_{a})^{2})}{4m^{2}}\right]^{1/2} 
\end{equation}
where $b = N, \Lambda, \Sigma, \textrm{or} \, \Xi$ and $a = \pi, K, \bar{K}, \textrm{or} \, \eta$.

We can repeat the calculations above for the $\Sigma$ multiplet. The mass matrix for $\Sigma$, analagous to \eq{massmatrix} is
\begin{equation} \label{massmatrixsigma}
\begin{bmatrix}
x_{\Sigma}  & y_{\Sigma} \\
y_{\Sigma} &  z_{\Sigma}
\end{bmatrix}
= \frac{m_{K}^{2}}{4 \pi f}
\begin{bmatrix}
- \frac{2}{3} b_{\mathcal{H}} - 2 \sigma_{P} & \quad  - \frac{2}{\sqrt{3}} b_{PO}  \\
 - \frac{2}{\sqrt{3}} b_{PO}  & \quad - \bar{\Delta}_{PO} - 2 \sigma_{P}
\end{bmatrix}.
\end{equation}
By diagonalizing the mass matrix, the masses of the physical states are obtained: 
\begin{align}
m_{\Sigma_{1}} &=  \left( m_{P}-2 \sigma_{P} \frac{m_{K}^{2}}{4 \pi f}\right) - \frac{1}{3} b_{\mathcal{H}} \left(1 + \sec{2 \theta_{\Sigma}}\right) - \frac{1}{2} \bar{\Delta}_{PO} \left(1 - \sec{2 \theta_{\Sigma}}\right) \\
m_{\Sigma_{2}} &=  \left( m_{P}-2 \sigma_{P} \frac{m_{K}^{2}}{4 \pi f}\right)- \frac{1}{3} b_{\mathcal{H}} \left(1 - \sec{2 \theta_{\Sigma}}\right) - \frac{1}{2} \bar{\Delta}_{PO} \left(1 + \sec{2 \theta_{\Sigma}}\right).
\end{align}
The mixing angle $\theta_{\Sigma}$ is between $-45^{\circ}$ and $45^{\circ}$ and is defined by
\begin{equation}
\tan{2 \theta_{\Sigma}}=\frac{ - \frac{2}{\sqrt{3}} b_{PO} }{ \frac{1}{3} b_{\mathcal{H}} - \frac{1}{2} \bar{\Delta}_{PO}}.
\end{equation}
The partial widths for $\Sigma_{1}$ and $\Sigma_{2}$ into $N \bar{K}$ are 
\begin{align} \label{sigmawidth}
\Gamma_{\Sigma_{1} N \bar{K}} &=  \left(\frac{1}{\sqrt{3}} \mathcal{C}_{PB} \cos{\theta_{\Sigma}} -\left(D_{OB}+F_{OB}\right) \sin{\theta_{\Sigma}} \right)^{2} \left( \frac{m_{N}}{m_{\Sigma_{1}}} \right) \frac{k_{N \bar{K}}^{3}(m_{\Sigma_{1}})}{2 \pi f^2} \\
\Gamma_{\Sigma_{2} N \bar{K}} &=  \left(+\frac{1}{\sqrt{3}} \mathcal{C}_{PB} \sin{\theta_{\Sigma}} + \left(D_{OB}+F_{OB}\right) \cos{\theta_{\Sigma}} \right)^{2} \left( \frac{m_{N}}{m_{\Sigma_{2}}}\right)\frac{k_{N \bar{K}}^{3}(m_{\Sigma_{2}})}{2 \pi f^2}.
\end{align}

\section{Discussion}\label{disc}

In the previous section, masses and partial widths were determined at leading order in terms of the 6 mass parameters and 3 couplings appearing in the LO chiral Lagrangian. No model-dependent assumptions were made. Although there are many parameters in all, a sufficient number of predictions, and comparisons with data, can constrain them. As such, this approach becomes more useful as more data becomes available. It also provides a general framework to understand data, relate models, and highlight unexpected parameter values determined from data that require microscopic explanation.

The results of the HB$\chi$PT calculations for all eight isospin multiplets are collected in table \ref{fulltable}. For simplicity, we only include the partial widths to $N \pi, N K, \textrm{or} \, N \bar{K}$. In principle, other partial widths could be included in the fit of the parameters. In subsection \ref{massrelations}, the two mass relations originally obtained by Diakonov and Petrov in \cite{Diakonov} are reproduced. The motivation for mixing is briefly reconsidered. In subsection \ref{diquark}, the parameter limit leading to the JW model is determined. In subsection \ref{Roper} Glozman's argument in \cite{Glozman} against the exotic nature of the $N(1440)$ is summarized, and more evidence casting doubt on a {\it purely} exotic $N(1440)$ is introduced. In subsection \ref{constraints}, GMO-type relations satisfied by the reduced partial widths, generalizing the inequalities by Cohen in \cite{Cohen}, are developed. The implications of the constraints for the mixing angle and the parameters in the chiral Lagrangian are discussed.

\begin{table}[b] 
\caption{\label{fulltable} Masses and reduced partial widths of anti-decuplet and octet pentaquarks in HB$\chi$PT. All masses are given relative to $m_{P}- 2\sigma_{P} \frac{m_{K}^{2}}{4 \pi f} $ in units of $\frac{m_{K}^{2}}{4 \pi f}$. Reduced partial widths are for decays to $N \pi$, $N K$, or $N \bar{K}$, at most one of which is allowed in each case.} 
\begin{center}
\begin{tabular}{|c|c|c|}
\hline
Particle & Mass & Reduced partial width\\
\hline
$\Theta$ &  $-2 b_{\mathcal{H}}$ & $2 \mathcal{C}_{PB}^{2}$ \\
\hline
$N_{1}$ & $ {\scriptstyle - \frac{2}{3} b_{\mathcal{H}} \left(1 + \sec{2 \theta_{N}}\right) +  \left(- b_O^{D} +  b_O^{F} - \frac{1}{2} \bar{\Delta}_{PO}\right)  \left(1 - \sec{2 \theta_{N}}\right)} $ & $\frac{3}{2} \left(\frac{1}{\sqrt{3}} \mathcal{C}_{PB} \cos{\theta_{N}} - \left(D_{OB}-F_{OB}\right) \sin{\theta_{N}} \right)^{2}$ \\ 
\hline
$N_{2}$ & ${\scriptstyle - \frac{2}{3} b_{\mathcal{H}} \left(1 - \sec{2 \theta_{N}}\right) +  \left(- b_O^{D} +  b_O^{F} - \frac{1}{2} \bar{\Delta}_{PO}\right)  \left(1 + \sec{2 \theta_{N}} \right)}$ & $ \frac{3}{2} \left(\frac{1}{\sqrt{3}} \mathcal{C}_{PB} \sin{\theta_{N}} + \left(D_{OB}-F_{OB}\right) \cos{\theta_{N}} \right)^{2}$ \\
\hline
$\Sigma_{1}$ & $- \frac{1}{3} b_{\mathcal{H}} \left(1 + \sec{2 \theta_{\Sigma}}\right) - \frac{1}{2} \bar{\Delta}_{PO} \left(1 - \sec{2 \theta_{\Sigma}}\right) $ & $ \left(\frac{1}{\sqrt{3}} \mathcal{C}_{PB} \cos{\theta_{\Sigma}} - \left(D_{OB}+F_{OB}\right) \sin{\theta_{\Sigma}} \right)^{2}$ \\
\hline
$\Sigma_{2}$ &$- \frac{1}{3} b_{\mathcal{H}} \left(1 - \sec{2 \theta_{\Sigma}}\right) - \frac{1}{2} \bar{\Delta}_{PO} \left(1 + \sec{2 \theta_{\Sigma}}\right)$& $ \left(\frac{1}{\sqrt{3}} \mathcal{C}_{PB} \sin{\theta_{\Sigma}} + \left(D_{OB}+F_{OB}\right) \cos{\theta_{\Sigma}} \right)^{2}$ \\
\hline
$\Lambda$ & $ - \frac{8}{3} b_{O}^{D} - \bar{\Delta}_{PO}$ & $ \frac{1}{3} \left(D_{OB} + 3 F_{OB}\right)^{2} $\\
\hline
$\Xi$ & $- 2 b_{O}^{D} - 2 b_{O}^{F} - \bar{\Delta}_{PO} $ & $ - $\\
\hline
$\Phi$ & $0$ & $ - $ \\
\hline
\end{tabular}
\end{center}
\end{table}

 \subsection{Relations between masses} \label{massrelations}
 
With six parameters and the masses of eight isospin multiplets, one can predict two relations analagous to the Gell-Mann-Okubo (GMO) relations \cite{Diakonov}:
\begin{align} 
\label{gmo1} 2 m_{N_{1}} + 2 m_{N_{2}} + 2 m_{\Xi} & = m_{\Sigma_{1}} + m_{\Sigma_{2}} + 3 m_{\Lambda}+ m_{\Theta} \\
2 m_{N_{1}} \left(1- \cos{2 \theta_{N}} \right) & + 2 m_{N_{2}} \left(1+ \cos{2 \theta_{N}} \right) + 4 m_{\Xi} \nn \\  \label{gmo2}
& = m_{\Sigma_{1}} \left(1- \cos{2 \theta_{\Sigma}} \right) +  m_{\Sigma_{2}} \left(1+ \cos{2 \theta_{\Sigma}} \right) + 6 m_{\Lambda}
\end{align}
Since \eq{gmo1} and \eq{gmo2} are consequences of flavor symmetry alone, no further assumption of the quark content of the flavor multiplets is required. The scenario considered in \cite{Diakonov} involved mixing between an anti-decuplet and an octet. The authors' phenomenological motivation for the scenario was the observation that the lowest excited ${\frac{1}{2}}^{+}$ baryons, the so-called Roper resonances, satisfy the GMO to a lesser degree than the ground state ${\frac{1}{2}}^{+}$ baryons and the next excited state ${\frac{1}{2}}^{+}$ baryons. Using the method of least squares, one finds that the error of the best fit is $24$ MeV for the Roper states compared to $3$ MeV for the ground state octet and $5$ MeV for the next excited octet.

The errors in the fit quoted above do not take into account the uncertainty associated with each of the mass measurements. The importance of the uncertainties is made apparent in \cite{Diakonov} though. As masses are varied within the uncertainty intervals, there are siginificant changes in the best fit. Surprisingly, in each of the two examples considered in \cite{Diakonov}, the alternative assignment of masses within the uncertainty intervals satisfies the usual {\it octet} GMO to a good degree. For example, the error of the fit to the octet GMO for $m_{N}=1470$, $m_{\Lambda}=1570$, $m_{\Sigma}=1635$, and $m_{\Xi}=1700$ is less than $2$ MeV. A satisfactory fit is obtained with no mixing at all thus taking away the phenomenological motivation for considering mixing in the first place. In other words, flavor symmetry and the imprecise mass data alone do not provide sufficient motivation for a particular mixing angle, or even a nonzero mixing angle. The large parameter space allows too much freedom

To further clarify the issue, one can use a $\chi^{2}$ test, which takes into account the uncertainty in the mass measurements. Minimizing $\chi^{2}$ with respect to three paramters when there are only four data points leaves only one degree of freedom. Although the result should be taken with a grain of salt, the Ropers satisfy the octet GMO to the expected degree of accuracy: $\chi^{2} = .85$.

\subsection{Parameters of the diquark model} \label{diquark}

One limitation of HB$\chi$PT is that it neither employs all that has been learned about low-energy QCD, nor does it aim for a microscopic explanation of the data. This is the domain of models. Jaffe and Wilczek developed a diquark model with a color-spin interaction Hamiltonian for the quarks that is motivated by one-gluon exchange in QCD. It has support in various parts of strong-interaction physics \cite{Jaffe}. The parameter space of the JW model is smaller than that of HB$\chi$PT by 3 parameters, and thus constrains the predicted mass spectrum to a greater degree. Five predictions can be made about the masses of the eight isospin multiplets. The mass predictions of the JW model from \cite{Jaffe} are shown in table \ref{dict}a. In addition, the model predicts ideal mixing, $\theta_{N}= 35.3^{\circ}$ and $\theta_{\Sigma}= -35.3^{\circ}$.

\begin{table}[b]
\caption{\label{dict} a) Masses predicted by JW model in \cite{Jaffe}. b) HB$\chi$PT limit that corresponds to the JW model.}
\begin{center}
a) \quad \begin{tabular}{|c|c|}
\hline
Particle & Mass \\
\hline 
$\Theta$ & $M_{0} + \mu$ \\
\hline 
$N_{1}$ & $M_{0} + 2 \mu + \alpha$ \\
\hline 
$N_{2}$ & $M_{0}$ \\
\hline 
$\Sigma_{1}$ & $M_{0} + \mu + \alpha$\\
\hline 
$\Sigma_{2}$ & $M_{0} + 3 \mu + 2 \alpha$\\
\hline 
$\Lambda$ & $M_{0} + \mu + \alpha$\\
\hline 
$\Xi$& $M_{0} + 2 \mu + 2 \alpha$ \\
\hline 
$\Phi$& $M_{0} + 2 \mu + 2 \alpha$ \\
\hline
\end{tabular}
\hspace{2cm}
b) \quad \begin{tabular}{|c|c|}
\hline
HB$\chi$PT & JW \\
\hline
$m_{P} - 2 \sigma_{P} \frac{m_{K}^{2}}{4 \pi f}$ & $M_{0} + 2 \mu + 2 \alpha$ \\
\hline
$\bar{\Delta}_{PO}  \frac{m_{K}^{2}}{4 \pi f}$ & $\frac{1}{3} \left(\alpha - \mu\right)$ \\ 
\hline
$b_{\mathcal{H}}  \frac{m_{K}^{2}}{4 \pi f}$ & $\frac{1}{2} \left(2 \alpha + \mu \right)$ \\
\hline
$b^{D}_{O}  \frac{m_{K}^{2}}{4 \pi f}$ & $\frac{1}{4} \left(\alpha + 2 \mu\right)$ \\
\hline
$b^{F}_{O}  \frac{m_{K}^{2}}{4 \pi f}$ & $-\frac{1}{12} \left(5 \alpha +  4 \mu \right)$ \\
\hline
$b_{PO}  \frac{m_{K}^{2}}{4 \pi f}$ & $\frac{1}{\sqrt{6}} \left(\alpha + 2 \mu \right) $ \\
\hline
\end{tabular}
\end{center}
\end{table}

The masses calculated in HB$\chi$PT in table \ref{fulltable} can be related to the predictions of quark models. For concreteness, we work with the JW model but the approach can be applied to other models for which there are a sufficient number of predictions to match on to. Table \ref{dict}b describes the limit of the mass parameters in HB$\chi$PT that corresponds to the JW model. With the right choice of only three parameters, the JW model can approximately fit the masses of $\Theta(1540)$, $N(1440)$, $N(1710)$, $\Sigma(1660)$, $\Xi(1860)$, and $\Phi(1860)$. However, it has difficulty fitting their partial widths \cite{Cohen, Jaffe}. In particular, the order magnitude difference in the reduced partial widths for $N(1440) \to N \pi$ and $\Theta \to N K$ cannot be accommodated. 

Two alternatives present themselves. If the identification of physical resonances with pentaquark states  is correct, it might be possible to adjust the JW model in the larger parameter space of HB$\chi$PT to fit the partial width data. For example, it might be possible to adjust the mixing angle to accommodate the partial widths and adjust the other parameters to maintain the fit of the masses. We pursue this in subsection \ref{constraints}. If this were possible, it might suggest that the quark model adopted requires some adjustment. On the other hand, the assignment of $N(1440)$ to the pentaquark multiplet might simply be incorrect. Evidence supporting this alternative is presented in subsection \ref{Roper}.

 \subsection{Constraining parameter space} \label{constraints}

We now turn to the LO reduced partial widths in table \ref{fulltable}. The first three widths, $\bar{\Gamma}_{\Theta N K}, \bar{\Gamma}_{N_{2} N \pi}, \textrm{and} \, \bar{\Gamma}_{N_{1} N \pi}$ are expressed in terms of two additional independent couplings, $\mathcal{C_{PB}}$ and $(D_{OB} + F_{OB})$. Thus, there is one relation satisfied by the three widths. A similar argument implies a relation satisfied by  $\bar{\Gamma}_{\Theta N K}, \bar{\Gamma}_{\Sigma_{2} N \bar{K}}, \textrm{and} \, \bar{\Gamma}_{\Sigma_{1} N \bar{K}}$. The two relations are
\begin{align}
\label{nwidthrel} 4 \tan^2{\theta_{N}} \, \bar{\Gamma}_{\Theta N K} & \left( \bar{\Gamma}_{N_{2} N \pi} + \bar{\Gamma}_{N_{1} N \pi}  - \frac{1}{4} \, \bar{\Gamma}_{\Theta N K} \right) \nn \\
  & = \left( 2 \tan^2{\theta_{N}} \, \bar{\Gamma}_{N_{2} N \pi} - 2 \, \bar{\Gamma}_{N_{1} N \pi}  +  \frac{1 -  \tan^2{\theta_{N}} }{2} \, \bar{\Gamma}_{\Theta N K} \right)^{2} \\
\label{sigwidthrel} \frac{8}{3} \tan^2{\theta_{\Sigma}} \, \bar{\Gamma}_{\Theta N K}  & \left( \bar{\Gamma}_{\Sigma_{2} N \bar{K}} + \bar{\Gamma}_{\Sigma_{1} N \bar{K}}  - \frac{1}{6} \, \bar{\Gamma}_{\Theta N K} \right) \nn \\
 & = \left( 2 \tan^2{\theta_{\Sigma}} \, \bar{\Gamma}_{\Sigma_{2} N \bar{K}} - 2 \, \bar{\Gamma}_{\Sigma_{1} N \bar{K}}  +  \frac{1 -  \tan^2{\theta_{\Sigma}} }{3} \, \bar{\Gamma}_{\Theta N K} \right)^{2}
\end{align}
Here $\bar{\Gamma}$ denotes a reduced partial width, i.e. a partial width without the phase space factor, such as $\bar{\Gamma}_{\Theta N K} = 2 \mathcal{C}_{P B}^{2}$. \Eq{nwidthrel} is simplified by dividing by $\bar{\Gamma}_{N_{2} N \pi}^{2}$. It then takes the form
\begin{equation}
\label{nwidthrelxy} 4 \tan^2{\theta_{N}} \, x  \left( 1 + y  - \frac{1}{4} \, x \right) = \left( 2 \tan^2{\theta_{N}} \, - 2 \, y  +  \frac{1 -  \tan^2{\theta_{N}} }{2} \, x \right)^{2}
\end{equation} 
 where $x= \bar{\Gamma}_{\Theta N K}/\bar{\Gamma}_{N_{2} N \pi}$ and $y=\bar{\Gamma}_{N_{1} N \pi} / \bar{\Gamma}_{N_{2} N \pi}$. \Eq{sigwidthrel} for the $\Sigma$ multiplets can also be simplified. It is not discussed further because of the lack of an established candidate for $\Sigma_{2}$.

\begin{figure}[htdp]
\caption{ \label{constr} The curve is the constraint for the relative reduced partial widths of the pentaquarks at leading order in HB$\chi$PT. Each plot is labeled by the choice of mixing angle $\theta_{N}$. The shaded region is the weaker constraint if NLO effects are large.}
\begin{center}
\includegraphics{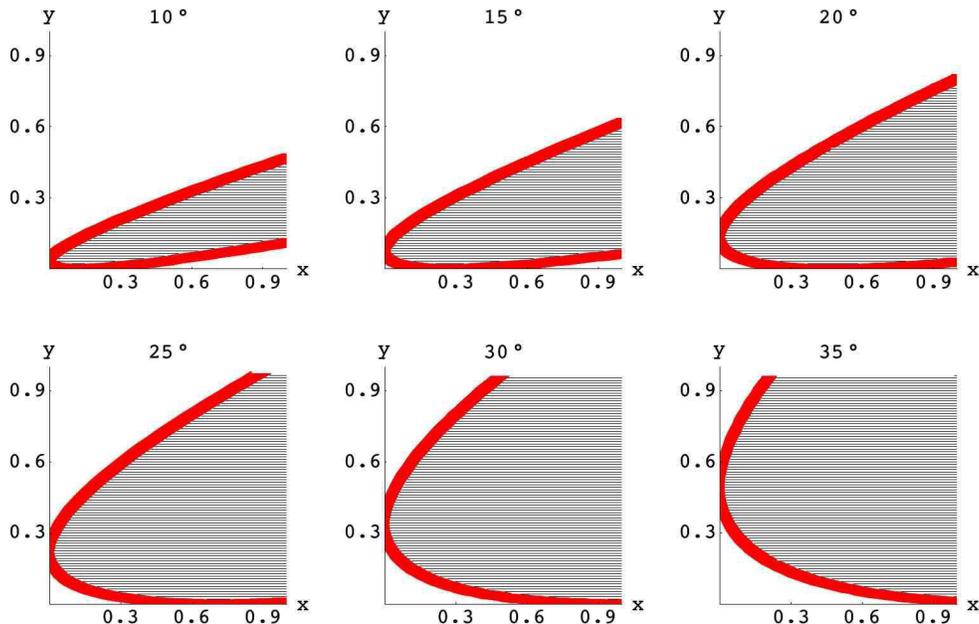}
\end{center}
\end{figure}

\begin{figure}[htdp]
\caption{ \label{idealgraph} a) Constraint curve and region for ideal mixing angle $\theta_{N}= 35.3^{\circ}$. The dot and the rectangle are the experimentally favored value and the uncertainty respectively.  b) Constraint curve and region for mixing angle $\theta_{N}= 12.2^{\circ}$. The dot and the rectangle are the experimentally favored value and the uncertainty respectively.}
\begin{center}
\includegraphics{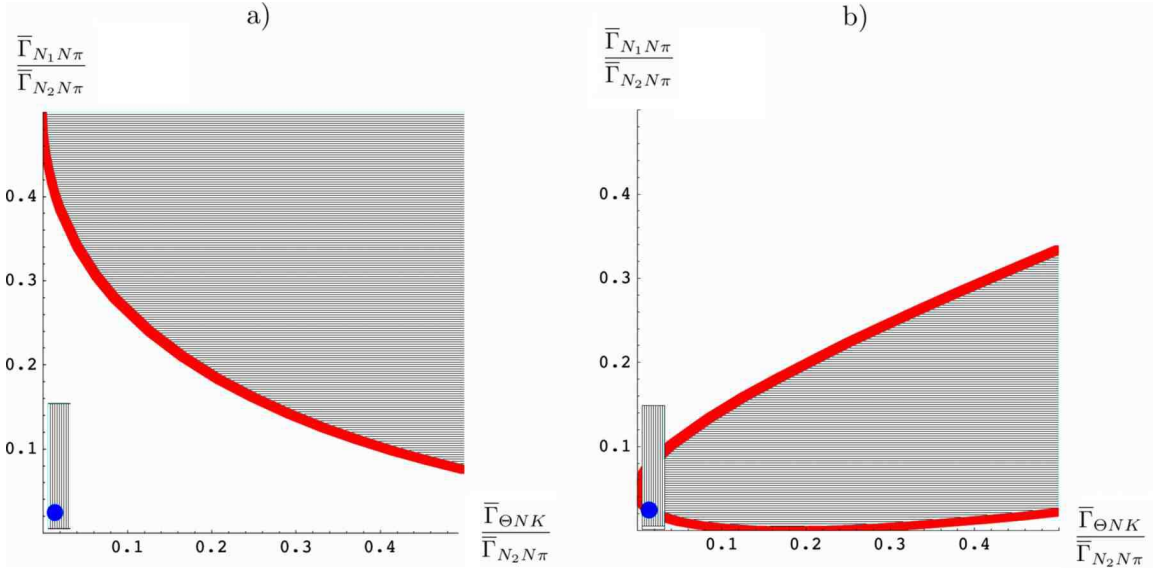}
\end{center}
\end{figure}

The data in table \ref{nwidthdata} from the PDG \cite{PDG} determines the experimentally favored values and uncertainties for $x$ and $y$: 
\begin{equation}\label{favored}
x = .014 \pm ^{.018}_{.008} \quad \textrm{and} \quad y=.024 \pm ^{.130}_{.019}.
\end{equation}
\begin{table}[b]
\caption{\label{nwidthdata} Width data for $\Theta(1540)$, $N(1440)$, and $N(1710)$. Mass, phase space, and total width are in MeV. The last two columns refer to decays to $N \pi$ or $N K$. }
\begin{center}
\begin{tabular}{|c|c|c|c|c|c|}
\hline
Particle & Mass & Phase space & Total width & Branching fraction & Reduced partial width \\ 
\hline
$\Theta(1540)$ & $1539.2 \pm 1.6$ &$107 \pm 3 $& $0.9 \pm 0.3 $& $1$ &   $.008\pm.003$  \\ 
\hline
$N(1440)$ & $1440 \pm^{30}_{10}$ & $379 \pm ^{58}_{19}$ & $350 \pm 100$ & $.65 \pm .05$& $.60 \pm .27 $\\
\hline
$N(1710)$ & $1710 \pm 30$ & $1029 \pm 87$ & $100 \pm^{150}_{50}$ &$ .15 \pm .05$ & $.015 \pm ^{.038}_{.010} $ \\
\hline
\end{tabular} \\
\end{center}
\end{table}
Figure \ref{constr} shows the curve defined by \eq{nwidthrelxy} for various mixing angles, including ideal mixing.
Figure \ref{idealgraph}a shows just the curve for ideal mixing along with the favored value of $(x,y)$ in \eq{favored}.
The figure clearly shows that ideal mixing $\theta_{N} = 35^{\circ}$ is inconsistent with the order magnitude difference in the reduced partial widths for $N(1440) \to N \pi$ and $\Theta \to N K$ in agreement with Cohen. However, the graphs in figure \ref{constr} suggest that a smaller mixing angle is consistent with the partial width data. We find that for mixing angles of $\theta_{N}= 5.5^{\circ}$ or $\theta_{N}= 12.2^{\circ}$, the experimentally favored value lies on the constraint curve as shown in figure \ref{idealgraph}b. The uncertainty in the favored value implies that the data is consistent with $0^{\circ} \lesssim \theta_{N} \lesssim 25^{\circ}$.

\begin{figure}[htdp]
\caption{ \label{imagloop} NLO correction of the pentaquark decay amplitude. Double lines are pentaquarks, single thin lines are nucleons, dashed lines are pions.}
\begin{center}
\includegraphics{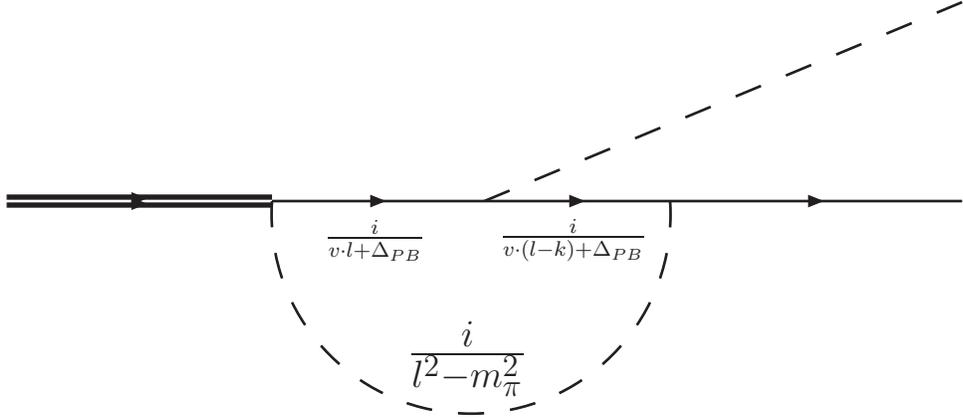}
\end{center}
\end{figure}

The relations in \eq{nwidthrel} and \eq{sigwidthrel} are possible because amplitudes are real at leading order and partial widths are proportional to the square of the amplitude. At NLO, the amplitudes pick up imaginary parts from loop diagrams. The square of the absolute magnitudes of the amplitudes then appear in the reduced partial widths. If the imaginary parts are small, the relations in \eq{nwidthrel} and  \eq{sigwidthrel} remain approximately satisfied. However, if the imaginary parts of the amplitudes at NLO are large enough, the relations above are not satisfied very well. This could happen because the mass difference $\Delta_{PB}$ appears in propagators in loop diagrams such as in figure \ref{imagloop}.
Such diagrams contribute terms proportional to $(\frac{\Delta_{PB}}{4 \pi f})^{2}$. This potential problem with loops in the chiral expansion for pentaquarks was side-stepped in \cite{Mohta} by an additional small coupling expansion. The addition of a pentaquark octet no longer allows such an expansion. If the imaginary parts of the decay amplitudes do become large at NLO, the best one can do is an inequality:
 \begin{equation}
\label{nwidthrelgeq} 4 \tan^2{\theta_{N}} \, x  \left( 1 + y  - \frac{1}{4} \, x \right) \geq \left( 2 \tan^2{\theta_{N}} \, - 2 \, y  +  \frac{1 -  \tan^2{\theta_{N}} }{2} \, x \right)^{2}
\end{equation}
This inequality was first considered by Cohen in the special case of ideal mixing or $\theta_{N}=35.3^{\circ}$ \cite{Cohen}. The inequality constrains the ratios of reduced partial widths only to the shaded region in figure \ref{constr}, rather than to the bounding curve. 

Two natural questions remain unanswered at this point. Are there values of the mass parameters that are consistent with the partial width data (in the sense discussed above) but that still allow the JW identification of resonances? Are there corresponding values of the couplings that fit the partial widths?

Since there are many parameters in HB$\chi$PT, a sufficient amount of data is required to pin them down. With only the six masses in the JW identification, the mass parameters, for example in column $\chi$PT 2 of table \ref{massparam}, can fit them exactly.  However, the partial widths of the resonances impose an additional constraint as described earlier. In the constrained parameter space, the best fit to the six masses appears in column $\chi$PT 1. For reference, the first two columns have parameter values in the  restricted part of the parameter space corresponding to the JW model. The first column uses values for the parameters derived, using table \ref{dict}, from those proposed in \cite{Jaffe} and motivated by the rest of the baryon spectrum. The second column has the best fit restricted to the parameter space of the JW model.
\begin{table}[t]
\caption{\label{massparam} Values of the HB$\chi$PT mass parameters in four approaches. JW 1) Fit assuming JW model with parameters motivated by baryon spectrum. JW 2) Fit assuming JW model with minimum $\chi^{2}$. $\chi$PT 1) Fit with minimum $\chi^{2}$ given mixing angle constraint. $\chi$PT 2) Fit with minimum $\chi^{2}$.}
\begin{center}
\begin{tabular}{|c|c|c|c|c|}
\hline
Parameter & JW 1 & JW 2 & $\chi$PT 1& $\chi$PT 2 \\ 
\hline
$m_{P} - 2 \sigma_{P} \frac{m_{K}^{2}}{4 \pi f}$ &$1850$&$1860$&$1860$&$1860$\\ 
\hline
$\bar{\Delta}_{PO} \frac{m_{K}^{2}}{4 \pi f}$&$0$&$9$&$99$&$60$ \\
\hline
$b_{\mathcal{H}}$ &$1.01$&$1.07$&$1.07$&$1.07$\\
\hline
$b^{D}_{O}$ &$0.50$&$0.49$&$0.30$&$0.40$\\
\hline
$b^{F}_{O}$ &$-0.50$&$-0.52$&$-0.63$&$-0.60$\\
\hline
$b_{PO}$ &$0.82$&$0.80$&$0.57$&$0.67$\\
\hline
\hline
$\theta_{N}$&$35.3$&$35.3$&$25.0$&$29.0$\\
\hline
$\chi^{2}$&$91.5$&$1.71$&$0.37$&$0$\\
\hline
\end{tabular} \\
\end{center}
\end{table}

Only the third column of table \ref{massparam} satisfies the mixing angle constraint imposed by partial width data and \eq{nwidthrelxy}. If the couplings are $\mathcal{C}_{PB} =0.07, D_{OB}=-0.42, \textrm{and} \, F_{OB}=0.11$ (or all of their negatives), the partial widths for $\Theta, N_{1}, N_{2}, \textrm{and} \, \Sigma$ are $1, 51, 140, \textrm{and} \, 20$ MeV, which are roughly right. As expected, the partial widths for $N_{1}$ and $N_{2}$ have to be at opposite ends of their uncertainty intervals to obtain a decent fit. 

As mentioned earlier, the minimizing of $\chi^{2}$ with such a limited data set should be taken with a grain of salt. Table \ref{massparam} only shows that the masses {\it and} partial widths in the JW identification can be fit by a mixed octet and anti-decuplet in HB$\chi$PT, with parameters corresponding to a small adjustment of the JW model. As more data becomes available, a more comprehensive exploration of the parameter space should be fruitful.

\subsection{Identity of the Roper} \label{Roper}

One of the arguments for suspecting that $N(1440)$ and its partners are unusual is that their masses do not fit neatly into quark models that have been successful in fitting much of the hadronic spectrum. An alternative to the JW identification of $N(1440)$ has been suggested by Glozman \cite{Glozman}. It is argued that $N(1440)$, $\Lambda(1600)$, $\Sigma(1660)$, and $\Delta(1600)$ have similar broad widths and could fill out a $\mathbf{56}$ of spin-flavor $SU(6)$. In addition, a constituent quark model with Goldstone boson exchange that has been successfully tested on the baryon spectrum explains their masses well \cite{Glozman2}.

There is further evidence for a relationship between $N(1440)$ and $\Delta(1600)$ that, to the best of our knowledge, has not been noted. The decay $\Delta(1600)$ to $N(1440) \pi$ has a large partial width given the available phase space. We will refer to these states as $\Delta^{*}$ and $N^{*}$. The ratio of the couplings $\mathcal{C}_{\Delta^{*}N^{*}}$ and $\mathcal{C}_{\Delta N}$ is related to the partial widths at LO in HB$\chi$PT by
\begin{equation}
{\left(\frac{\mathcal{C}_{\Delta^{*}N^{*}}}{\mathcal{C}_{\Delta N}}\right)}^{2} =\frac {\Gamma \left(\Delta^{*} \to N^{*} \pi \right)}{ \Gamma \left(\Delta \to N \pi \right)} \frac{m_{\Delta^{*}}}{m_{\Delta}} \frac{m_{N}}{m_{N^{*}}} \frac{k^{3}}{k_{*}^{3}}
\end{equation}
where $k_{*}$ and $k$ are the pion momenta in the respective processes. The greatest uncertainty in this quantity comes from the momentum $k_{*}$. The large uncertainties in the masses of $\Delta^{*}$ and $N^{*}$ reported in the PDG allow for a pion momentum from $0$ to $200$ MeV. There is an additional uncertainty in the partial width of $\Delta^{*}$. In all but the simultaneous ends of both uncertainty intervals, the ratio of the couplings is anywhere from one to much larger than one.

To pin down a more precise estimate that avoids combining of systematic errors, we extract all relevant quantities from individual experiments. As it turns out, there are only two experiments reporting the necessary data \cite{Manley, Vrana}. The PDG only includes the first of these in their fits \cite{PDG}.

The data appear in table \ref{deltadata}. The last column is the calculated value of $|\mathcal{C}_{\Delta^{*}N^{*}}|$, and uses $|\mathcal{C}_{\Delta N}| = 1.6$.
\begin{table}[b]
\caption{\label{deltadata} Data for $\Delta(1600)$ and $N(1440)$.  Momenta $k_{*}$ are calculated from masses and $^{\mathbf{P}}$ denotes P-wave-only data.}
\begin{center}
\begin{tabular}{|c|c|c|c|c|c|c|}
\hline
Source & $m_{\Delta^{*}}$ & $m_{N^{*}}$ & $k_{*}$ & $\Gamma_{\Delta^{*}}$ &$\Gamma_{\Delta^{*}N^{*}\pi}/ \Gamma_{\Delta^{*}}$& $ |\mathcal{C}_{\Delta^{*}N^{*}}| $ \\ 
\hline
Manley '92 & $1706 \pm 10$ &$1462 \pm 10 $&$187 \pm 16 $ & $430 \pm 73$ &${0.21^\mathbf{P}  \pm 0.05}$& $1.76 \pm .38$ \\ 
\hline
Vrana '00 & $1687 \pm 44$ &$1479 \pm 80$ &$ 146 \pm 110$ & $493 \pm 75$ &$ 0.13 \pm 0.04 $ &$ 1.47 \pm .86$\\
\hline
\end{tabular} \\
\end{center}
\end{table}

The size of the coupling suggests that the $\Delta(1600)$ and $N(1440)$ wavefunctions have significant overlap. The $\Delta(1600)$ cannot be a candidate for the spin $\frac{3}{2}$ pentaquark because its flavor quantum numbers are not those of a member of an anti-decuplet. It is possible that $\Delta(1600)$ and $N(1440)$ fill out a $\mathbf{56}$ of $SU(6)$ as proposed by Glozman. If this picture is right, it would take away some of the phenomenological motivation for considering mixing of the multiplets. However, it is also possible that the $N(1440)$ has a more complicated structure. It could have components of a radially excited nucleon and of both pentaquark multiplets as suggested in \cite{Maltman}. In order to explain the large coupling to $\Delta(1600)$, $N(1440)$ would have to have a substantial component of the radially excited nucleon. Finally, it could be that $N(1440)$ is actually two separate overlapping resonances. There is some data to support this claim \cite{Morsch}. It would allow the results in both \cite{Glozman} and \cite{Jaffe} to be correct, but about distinct resonances.

\section{Conclusion} \label{conc}

Flavor symmetry constraints on {\it partial widths} yield interesting results for pentaquark candidates. HB$\chi$PT provides a simple framework to analyze flavor symmetry constraints on partial widths and masses. Leading order calculations reproduce familiar flavor symmetry results such as GMO-type mass relations for mixed multiplets. In addition, the chiral expansion of decay amplitudes implies a new GMO-type relation for partial widths at leading order. Some care must be taken with the chiral expansion for decays of heavier pentaquarks where the pions are more energetic. As more data becomes available, a comprehensive analysis in HB$\chi$PT should prove fruitful.

\subsection{Acknowledgements}
I would like to thank Iain Stewart for many helpful discussions. I would also like to thank the JaxoDraw team for their great software \cite{JaxoDraw}. This work was supported in part by the NSF under the Mazur/Taubes research grant and by the DOE under cooperative research agreement DF-FC02-94ER40818.


\begin{thebibliography}{99}
\bibitem{Trilling}
For a recent overview of the experimental situation, see G. Trilling in \cite{PDG}.

%\cite{Mohta:2004yf}
\bibitem{Mohta}
V.~Mohta,
%``Pentaquark masses in chiral perturbation theory,''
[arXiv:hep-ph/0406233].
%%CITATION = HEP-PH 0406233;%%

%\cite{Praszalowicz:2004xh}
\bibitem{Pras}
M.~Praszalowicz,
%``SU(3) constraints on cryptoexotic pentaquarks,''
arXiv:hep-ph/0410086.
%%CITATION = HEP-PH 0410086;%%

%\cite{Jaffe:2003sg}
\bibitem{JaffeWilczek}
R.~L.~Jaffe and F.~Wilczek,
%``Diquarks and exotic spectroscopy,''
Phys.\ Rev.\ Lett.\  {\bf 91}, 232003 (2003)
[arXiv:hep-ph/0307341].
%%CITATION = HEP-PH 0307341;%%

%\cite{Karliner:2003dt}
\bibitem{KarlinerLipkin}
M.~Karliner and H.~J.~Lipkin,
%``A Diquark-Triquark Model for the KN Pentaquark,''
Phys.\ Lett.\ B {\bf 575}, 249 (2003)
[arXiv:hep-ph/0402260].
%%CITATION = HEP-PH 0402260;%%

%\cite{Diakonov:2003jj}
\bibitem{Diakonov}
D.~Diakonov and V.~Petrov,
%``Where are the missing members of the baryon antidecuplet?,''
Phys.\ Rev.\ D {\bf 69}, 094011 (2004)
[arXiv:hep-ph/0310212].
%%CITATION = HEP-PH 0310212;%%

%\cite{Jenkins:2004vb}
\bibitem{ManoharExotic}
E.~Jenkins and A.~V.~Manohar,
%``1/N(c) expansion for exotic baryons,''
JHEP {\bf 0406}, 039 (2004)
[arXiv:hep-ph/0402024].
%%CITATION = HEP-PH 0402024;%%

%\cite{Jaffe:2004at}
\bibitem{JaffeJain}
R.~L.~Jaffe and A.~Jain,
%``Implications of the present bound on the width of the Theta(1540)+,''
[arXiv:hep-ph/0408046].
%%CITATION = HEP-PH 0408046;%%

%\cite{Hong:2004xn}
\bibitem{Hong}
D.~K.~Hong, Y.~J.~Sohn and I.~Zahed,
%``A diquark chiral effective theory and exotic baryons,''
Phys.\ Lett.\ B {\bf 596}, 191 (2004)
[arXiv:hep-ph/0403205].
%%CITATION = HEP-PH 0403205;%%

%\cite{Cohen:2004gu}
\bibitem{Cohen}
T.~D.~Cohen,
%``Phenomenological constraints on the Jaffe-Wilczek model of pentaquarks,''
[arXiv:hep-ph/0402056].
%%CITATION = HEP-PH 0402056;%%

%\cite{Glozman:2003vb}
\bibitem{Glozman}
L.~Y.~Glozman,
%``Can low-lying Roper states be explained as antidecuplet members?,''
Phys.\ Rev.\ Lett.\  {\bf 92}, 239101 (2004)
[arXiv:hep-ph/0309092].
%%CITATION = HEP-PH 0309092;%%

%\cite{Glozman:1997ag}
\bibitem{Glozman2}
L.~Y.~Glozman, W.~Plessas, K.~Varga and R.~F.~Wagenbrunn,
%``Unified description of light- and strange-baryon spectra,''
Phys.\ Rev.\ D {\bf 58}, 094030 (1998)
[arXiv:hep-ph/9706507].
%%CITATION = HEP-PH 9706507;%%

%\cite{Eidelman:2004wy}
\bibitem{PDG}
S.~Eidelman {\it et al.}  [Particle Data Group Collaboration],
%``Review of particle physics,''
Phys.\ Lett.\ B {\bf 592}, 1 (2004).
%%CITATION = PHLTA,B592,1;%%

%\cite{Ko:2003xx}
\bibitem{Ko}
P.~Ko, J.~Lee, T.~Lee and J.~h.~Park,
%``Chiral perturbation theory for pentaquark baryons and its applications,''
arXiv:hep-ph/0312147.
%%CITATION = HEP-PH 0312147;%%

%\cite{Mehen:2004dy}
\bibitem{Mehen}
T.~Mehen and C.~Schat,
%``Determining pentaquark quantum numbers from strong decays,''
Phys.\ Lett.\ B {\bf 588}, 67 (2004)
[arXiv:hep-ph/0401107].
%%CITATION = HEP-PH 0401107;%%

%\cite{Grinstein:2004kn}
\bibitem{Grinstein}
B.~Grinstein and M.~A.~Savrov,
%``SU(3) decay amplitudes of pentaquarks into decouplet baryons,''
arXiv:hep-ph/0408346.
%%CITATION = HEP-PH 0408346;%%

%\cite{Golbeck:2004ix}
\bibitem{Golbeck}
S.~M.~Golbeck and M.~A.~Savrov,
%``Pentaquark decay amplitudes from SU(3) flavor symmetry,''
arXiv:hep-ph/0406060.
%%CITATION = HEP-PH 0406060;%%

%\cite{Jenkins:1990jv}
\bibitem{Jenkins}
E.~Jenkins and A.~V.~Manohar,
%``Baryon Chiral Perturbation Theory Using A Heavy Fermion Lagrangian,''
Phys.\ Lett.\ B {\bf 255}, 558 (1991).
%%CITATION = PHLTA,B255,558;%%

%\cite{Jaffe:2004ph}
\bibitem{Jaffe}
R.~L.~Jaffe,
%``Exotica,''
[arXiv:hep-ph/0409065].
%%CITATION = HEP-PH 0409065;%%

%\cite{Manley:1992yb}
\bibitem{Manley}
D.~M.~Manley and E.~M.~Saleski,
%``Multichannel resonance parametrization of pi N scattering amplitudes,''
Phys.\ Rev.\ D {\bf 45}, 4002 (1992).
%%CITATION = PHRVA,D45,4002;%%

%\cite{Vrana:1999nt}
\bibitem{Vrana}
T.~P.~Vrana, S.~A.~Dytman and T.~S.~H.~Lee,
%``Baryon resonance extraction from pi N data using a unitary multichannel
%model,''
Phys.\ Rept.\  {\bf 328}, 181 (2000)
[arXiv:nucl-th/9910012].
%%CITATION = NUCL-TH 9910012;%%

%\cite{Maltman:2004qd}
\bibitem{Maltman}
K.~Maltman,
%``Quark model perspectives on pentaquark exotics,''
arXiv:hep-ph/0408144.
%%CITATION = HEP-PH 0408144;%%

%\cite{Morsch:2000xi}
\bibitem{Morsch}
H.~P.~Morsch and P.~Zupranski,
%``Structure of the P(11) (1440 MeV) resonance from alpha p and pi N
%scattering,''
Phys.\ Rev.\ C {\bf 61}, 024002 (2000).
%%CITATION = PHRVA,C61,024002;%%

%\cite{Binosi:2003yf}
\bibitem{JaxoDraw}
D.~Binosi and L.~Theussl,
%``JaxoDraw: A graphical user interface for drawing Feynman diagrams,''
Comput.\ Phys.\ Commun.\  {\bf 161}, 76 (2004)
[arXiv:hep-ph/0309015].
%%CITATION = HEP-PH 0309015;%%
\end{thebibliography}
\end{document}